\documentclass[12pt]{article}

\usepackage{newtxtext,newtxmath}

\usepackage{graphicx}

\usepackage[letterpaper,margin=1in]{geometry}

\renewenvironment{abstract}
	{\quotation}
	{\endquotation}

\date{}

\makeatletter
\renewcommand{\fnum@figure}{\textbf{Figure \thefigure}}
\renewcommand{\fnum@table}{\textbf{Table \thetable}}
\makeatother

\usepackage{scicite}

\usepackage{url}

\newcommand{\ket}[1]{\mathinner{|{#1}\rangle}}
\newcommand{\bra}[1]{\mathinner{\langle{#1}|}}

\def\scititle{
	Hall conductivity reveals the nature of quantum coherence in strongly correlated metals
}
\title{\bfseries \boldmath \scititle}

\author{
	Emily~Z.~Zhang$^{1,2}$,
	Thomas~P.~Devereaux$^{1,2\ast}$\and
	\small$^{1}$Stanford Institute for Materials and Energy Sciences, SLAC National Accelerator Laboratory, Menlo Park, USA.\and\
    \small{$^{2}$
    Department of Materials Science and Engineering, Stanford University, Stanford, USA}\and\
	\small$^\ast$Corresponding author. Email: tpd@stanford.edu
}

\begin{document} 

\maketitle

\begin{abstract} \bfseries \boldmath
Linear-in-temperature resistivity is a hallmark for strange metallic transport, and appears universally in many strongly correlated electron systems. However, the focus on the longitudinal channel often overshadows the profound microscopic insights contained within the transverse response. Here, we utilize numerically exact determinantal quantum Monte Carlo simulations of the doped Hubbard model in a magnetic field to calculate longitudinal and transverse transport. We demonstrate that while the resistivity is robustly $T$-linear across parameter sets, the Hall response is highly sensitive to particle-hole asymmetry, Fermi surface topology, and many-body correlation effects. Specifically, the combination of these effects determine a crossover scale in which the system becomes quantum-coherent, and is reflected in the Hall conductivity. Our results demonstrate that while the $T$-linearity in resistivity appears universal, the Hall response reveals a crossover from semi-classical to quantum-coherent transport otherwise masked in the longitudinal channel. 
\end{abstract}

\noindent

Electrical transport provides one of the most direct probes of the excitations in quantum materials, yet it often remains the least understood. Conventional metals with well-defined quasiparticles obey Boltzmann transport theory, predicting a quadratic-in-temperature resistivity $\rho\sim T^2$ at low temperatures and resistivity saturation near the Mott-Ioffe-Regel (MIR) limit~\cite{gunnarsson_colloquium_2003-1}. In the case of strong electron correlations, however, this quasiparticle picture breaks down, leaving it unclear whether metallic transport principles survive beyond the Boltzmann regime. 

A canonical example of strange metallic transport in strongly correlated electron systems is in the high temperature phase of the cuprate superconductors~\cite{keimer_quantum_2015-1,phillips_stranger_2022-1}, where $\rho$ remains linear beyond the MIR limit~\cite{takagi_systematic_1992,martin_temperature_1988,gurvitch_resistivity_1987,barisic_universal_2013,ando_electronic_2004,hussey__universality_2004-1}, frequently accompanied by an inverse Hall angle $\cot(\theta_H)\sim T^2$~\cite{ono_strong_2007,hwang_scaling_1994,chien_effect_1991,ayres_incoherent_2021}. The coexistence of these temperature scalings motivated the influential proposal that longitudinal and transverse transport are governed by distinct relaxation lifetimes~\cite{anderson_hall_1991-1,ong_geometric_1991,nagaosa_normal-state_1990,lee_gauge_1992,coleman_how_1996-1}. However, both the microscopic mechanism behind these lifetimes and the universality of these temperature scalings in strongly correlated systems remain debated~\cite{stojkovic-acute_theory_1997,varma_phenomenology_1989-1,shastry_extremely_2018,hartnoll_scaling_2015-1}. Furthermore, experiments indicate that, while the linear resistivity is remarkably robust, the quadratic scaling of the inverse Hall angle can vary across materials~\cite{phillips_stranger_2022-1}, doping~\cite{ando_nonuniversal_1999}, oxygen concentration~\cite{jin_hall_1998}, and temperature~\cite{ono_strong_2007,hayes_superconductivity_2021,ando_evolution_2004,wu_antiferromagnetic_2026}. Understanding the nuances of the Hall response in these experiments therefore necessitates uncovering the origins behind its non-universality. 

Resolving these longstanding puzzles require determining whether the Hall conductivity is governed by the same processes that dictate resistivity. While longitudinal transport measures the net movement of charges through a material, transverse transport probes how carriers explore closed trajectories and respond to broken time-reversal symmetry. Consequently, the Hall conductivity is intrinsically more sensitive than resistivity to the topology of the Fermi surface, particle-hole asymmetry, and many-body correlations. Uncovering how this sensitivity shapes the Hall response is a crucial step in identifying the microscopic principles of transport beyond single-particle scattering.

The Hubbard model with next-nearest-neighbour hopping $t'$ provides a minimal microscopic platform for studying correlated electron systems in the absence of phonons or disorder~\cite{white_numerical_1989-1,loh_sign_1990}. Previous numerical studies using determinantal Quantum Monte Carlo (DQMC), a numerically exact method, demonstrated $T$-linear resistivity~\cite{huang_strange_2019-3} and identified a quadratic $\cot{\theta_H}$ scaling within the limit of $t'=0$ and $B\to0$~\cite{wang_dc_2020}. Here, we explore beyond these symmetric limits by computing longitudinal and transverse transport directly from current-current correlation functions using DQMC in the presence of a finite $B$ and $t'$ over a wide range of dopings and temperatures. This approach allows us to determine $\rho$, the Hall coefficient $R_H$, and $\cot{\theta_H}$, providing an unbiased microscopic characterization of transport in the doped Hubbard model. 

Across a wide parameter range, we identify a clear separation between longitudinal and transverse transport. The linear-in-$T$ resistivity remains robust at temperatures far beyond the MIR limit, exhibiting only weak sensitivity to $t'$, doping, and a coherence scale extracted from double occupancies that separates a semi-classical from a quantum regime. In contrast, the Hall conductivity changes qualitatively with $t'$, displaying distinct signs and temperature dependences.
Specifically, $t'$ affects the sign of $R_H$ not only through the Fermi surface geometry – as predicted from a Boltzmann transport picture \cite{ong_geometric_1991} – but also through the spectral weight redistribution of low lying states in the presence of interactions. The temperature dependence reflects the ability of charge carriers to execute coherent flux-enclosing hopping loops. Because such processes are absent from the leading longitudinal response, the Hall conductivity becomes a sensitive probe of the crossover from semi-classical transport to a quantum-coherent regime. Most importantly, the temperature at which $\sigma_{xy}$ departs from its high-temperature behavior coincides with the minimum in the average double occupancies, indicating that the Hall conductivity directly tracks the onset of many-body quantum coherence. While the longitudinal channel primarily reflects the presence of thermally active charge carriers, the Hall response is governed by flux-enclosing hopping processes and is therefore sensitive to the emergence of coherent electronic motion. Consequently, the inverse Hall angle exhibits qualitatively different integer power laws, whereas the linear resistivity remains largely unchanged. Our findings provide a framework for understanding the Hall response in the strange metal regime without invoking quasiparticles or phenomenological scattering rates.

\begin{figure}
    \centering
    \includegraphics[width=0.6\textwidth]{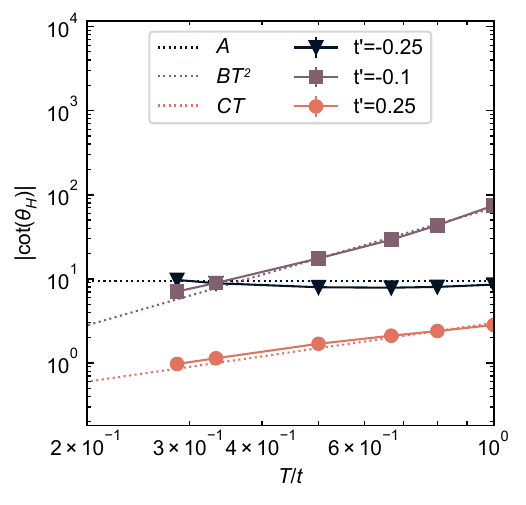}
    \caption{\textbf{Log-log plot of $\left|\cot(\theta_H)\right|$ as a function of temperature obtained from analytic continuation.}
        Power-law scalings are indicated with dotted lines as guides to the eye. Error bars are shown on the plot, but are smaller than the marker sizes. Parameters:  $U/t=6$, $\langle n\rangle=0.7$, $B=0.0625\Phi_0/a^2$. }
    \label{fig:cot_theta_H}
\end{figure}

Fig.~\ref{fig:cot_theta_H} shows the temperature dependence of the Hall angle, plotted as $\left|\cot(\theta_H)\right|$ on logarithmic scales for different $t'$ and fixed doping. Power-law curves $\propto T^\alpha$ are shown for $\alpha={0,1,2}$ as guides to the eye. Upon first glance, a striking feature of Fig.~\ref{fig:cot_theta_H} is that the temperature dependence is described by simple integer powers of $T$. Rather than exhibiting non-integer exponents characteristic of critical scaling, $\left|\cot(\theta_H)\right|$ being described by $T^0, T^1,$ or $T^2$ over extended temperature intervals suggests an expansion in inverse powers of temperature and motivates the analysis developed below.

For small $|t'|$, $\cot(\theta_H)$ exhibits an extended approximate $T^2$ regime, consistent with previous numerical studies on the Hubbard model~\cite{wang_dc_2020}. However, the scaling exponent $\alpha$ varies strongly with both $t'$, changing  $\left|\cot(\theta_H)\right|\sim T$ to $t'>0$ and $\left|\cot(\theta_H)\right|\sim \text{const}$ for $t'<0$. The Hall angle is also dependent on $U$, magnetic field, and doping (see Fig.~\ref{fig:U_nflux_doping_sweep} in the Supplementary Text), albeit the temperature dependence is not as dramatic as with varying $t'$. In other words, the quadratic scaling of $T$ occupies a narrow region in parameter space in the Hubbard-Hofstader model, rather than being ubiquitous to the strange metal regime. 

Since the Hall angle is determined by both the longitudinal and transverse transport coefficients through 
\begin{align}
\cot{\theta_H}=\frac{\rho_{xx}}{R_H},
\end{align}
this breakdown of the quadratic scaling may arise from $\rho_{xx}$, $R_H$, or an interplay between the two. We isolate these contributions by examining the longitudinal and Hall responses individually.

\begin{figure}
    \centering
    \includegraphics[width=0.9\textwidth]{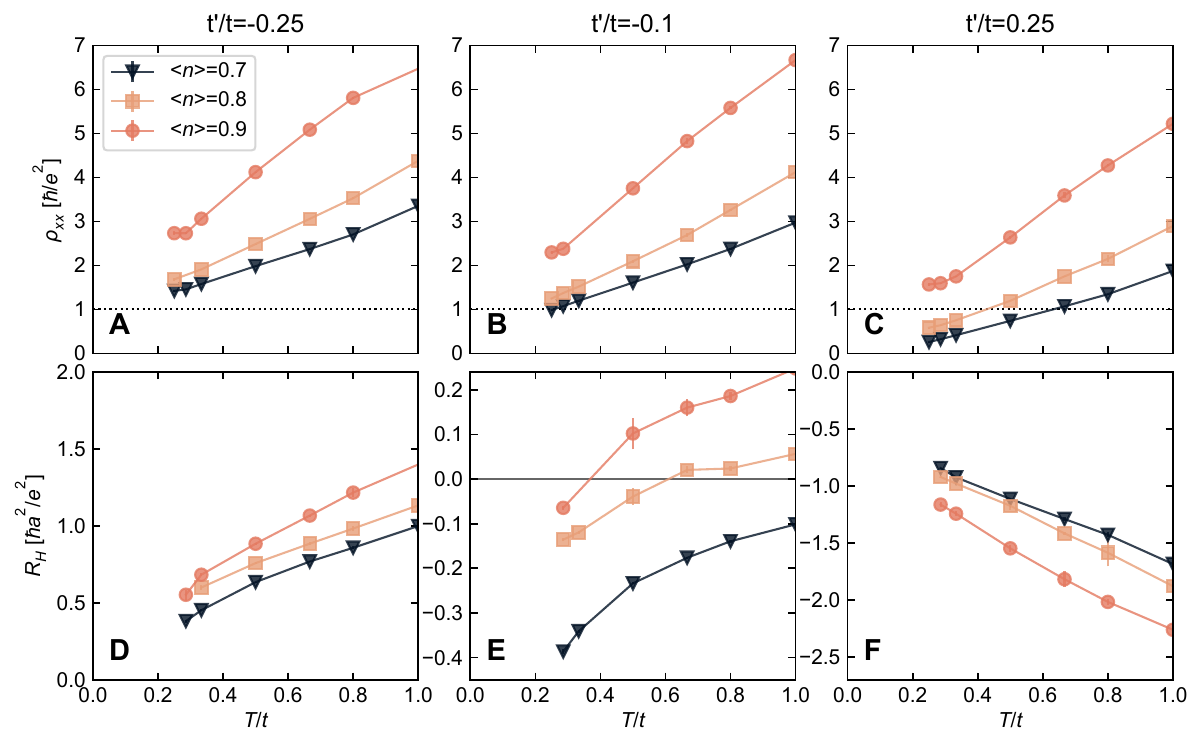}
    \caption{\textbf{Longitudinal resistivity $\rho_{xx}$ and Hall coefficient $R_H$ as a function of doping, $t'$, and temperature.}
        (\textbf{A})-(\textbf{C}) $\rho_{xx}$ vs $T$.
        The Mott-Ioffe-Regel criterion is marked by the black dotted line.
        (\textbf{D})-(\textbf{F}) $R_H$ vs $T$. 
        Parameters: $U/t=6$, $B=0.0625\Phi_0/a^2$. }
    \label{fig:rho_RH}
\end{figure}

Fig.~\ref{fig:rho_RH} compares the temperature dependence of $\rho_{xx}$ and $R_H$ across several dopings and values of $t'$ in the presence of a magnetic field. From Fig.~\ref{fig:rho_RH}(A)-(C), $\rho_{xx}$ remains approximately linear in $T$ across a wide range of dopings, demonstrating that transport in the longitudinal channel is considerably insensitive to changes in band structure. This robustness is consistent with previous numerical studies performed without a magnetic field, where linearity in $T$ persists far past the MIR limit~\cite{huang_strange_2019-3}. 

In contrast, the Hall coefficient in panels Fig.~\ref{fig:rho_RH}(D)-(F) displays qualitatively distinct behavior for $t'=-0.25$, $t'=-0.1$, $t'=0.25$, including variations in both its sign and temperature dependence. Only the case of $\langle n \rangle=0.7$ in panel (B) exhibits the $1/T$ scaling necessary for $\cot{\theta_H}\sim T^2$, accompanied by a negative sign. These results suggest that $R_H$ is highly dependent on microscopic details, especially band structure, and fundamentally governs the non-universal temperature scaling in $\cot(\theta_H)$. 

\begin{figure}
    \centering
    \includegraphics[width=0.9\textwidth]{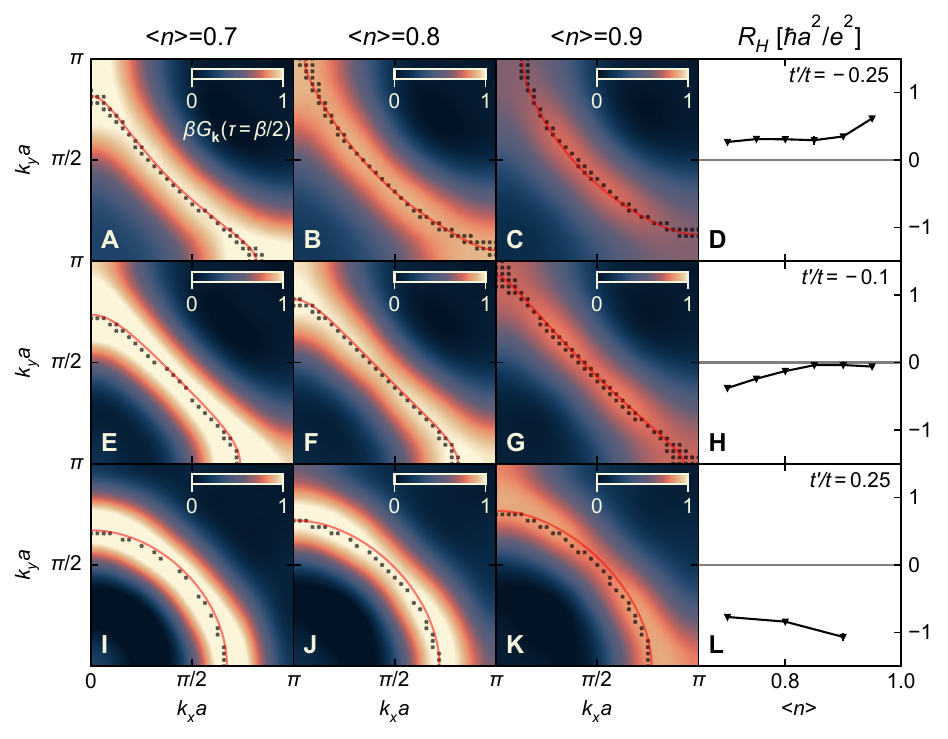}
    \caption{\textbf{Doping and $t'$ dependence of the spectral function proxy and Hall coefficient.} 
        The rows correspond to varying values of $t'$ and columns to $\beta G_\mathbf{k}(\tau=\beta/2)$ and $R_H$ versus doping. The non-interacting Fermi surface is denoted with a red line, and the black dots denote where $n_k=0.5\pm0.02$ in our data. A fine momentum grid was obtained using twisted boundary conditions for an 8x8 lattice, resulting in a 64x64 grid. Parameters:  $U/t=6$, $T/t=0.25$, $B=0.0625\Phi_0/a^2$. }
    \label{fig:spectralRH}
\end{figure}

The sensitivity of $R_H$ to doping and $t'$ points to two central issues: the microscopic origin of its sign, and the mechanism driving its temperature dependence. To address the first, we compute a proxy for the spectral function shown in Fig.~\ref{fig:spectralRH}, which illustrates how the Hall coefficient tracks the evolution of low-energy spectral weight through doping and band structure via $t'$. Because the notion of a Fermi surface becomes ambiguous in the strongly correlated regime, we examine the contour where $n_\mathbf{k}=0.5$ (marked by black dots) as a proxy. 

In general, the sign of $R_H$ tracks whether the Fermi surface closes around $(0,0)$ (electron-like) or $(\pi,\pi)$ (hole-like), consistent with semi-classical Boltzmann theory~\cite{ong_geometric_1991}. Notably, most panels broadly follow this expectation, even deep in the strongly correlated regime. 

However, Fig.~\ref{fig:spectralRH}(A) presents a notable exception. The non-interacting Fermi surface and the contour of $n_\mathbf{k}$ are both electron-like, yet the Hall coefficient remains positive. At the same time, the spectral weight near the antinodes $(\pi,0)$ is broadened and no longer follows the shape of the Fermi surface. This distribution of low-energy states appears more hole-like than suggested by the bare band geometry alone, signaling a breakdown of the conventional Boltzmann picture that depends only on a sharp quasiparticle Fermi surface. 

More broadly, interactions can smear and deform the low-energy spectra, leading to increasingly ambiguous boundaries between electron- and hole-like Fermi surfaces. These effects are especially pronounced for $t'<0$, where interaction-induced spectral broadening modifies the effective transport geometry. The Hall coefficient is therefore sensitive to the correlation-induced rearrangement of spectral weight, a feature unaccounted for in the geometrical Fermi surface picture~\cite{ong_geometric_1991}. 

\begin{figure}
    \centering
    \includegraphics[width=0.6\textwidth]{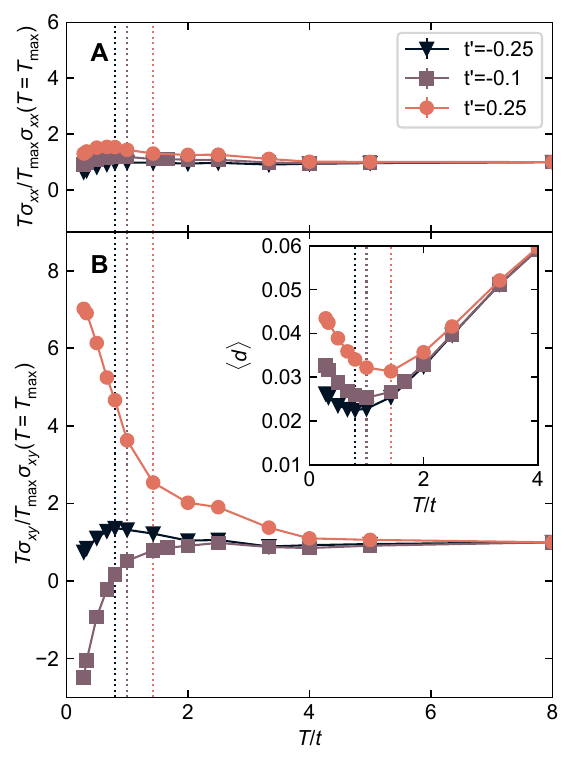}
    \caption{ \textbf{Longitudinal and Hall conductivity divided by high temperature asymptotic scaling $1/T$ versus temperature. }
        (\textbf{A}) Longitudinal conductivity $T\sigma_{xx}$ and (\textbf{B}) Hall conductivity $T\sigma_{xy}$ with an inset displaying the average double occupancy $\langle d \rangle=\langle n_\uparrow n_\downarrow \rangle$ versus temperature.  Each value of conductivity is normalized by $T_\text{max}\sigma(T=T_{\text{max}})$, where $T_{\text{max}}=8t$. $\sigma_{xx}$ was computed from MaxEnt, while $\sigma_{xy}$ from a Matsubara frequency proxy for $R_H$~\cite{methods}.  The dashed vertical lines mark the minima in $\langle d \rangle$ corresponding to each $t'$.
        Parameters: $U/t=6$, $B=0.0625\Phi_0/a^2$, $\langle n\rangle = 0.7$. 
    }
    \label{fig:Tsigma}
\end{figure}

To understand the temperature dependence in $R_H$, we examine the operator forms of its composite parts $\sigma_{xx}$ and $\sigma_{xy}$, related to $R_H$ through 
\begin{align}
    R_H=\frac{1}{B}\frac{\sigma_{xx}}{\sigma_{xx}^2+\sigma_{xy}^2}. 
\end{align}
Fig. \ref{fig:Tsigma} shows the temperature dependence of $T\sigma_{xx}$ and $T\sigma_{xy}$ for different values of $t'$. Here, $T\sigma_{xx}$ remains constant across a broad temperature range in Fig.~\ref{fig:Tsigma}A, contrasted by Fig.~\ref{fig:Tsigma}B for the Hall conductivity. At high temperatures ($T\gg U \gg t,t'$), all three curves approach a common asymptotic regime where $T\sigma_{xy}\sim\text{const.}$ As the temperature is reduced, $\sigma_{xy}$ diverges from its high-temperature behavior at different temperatures for different $t'$. Specifically, the $t'=0.25$ curve breaks away first, followed by $t'=-0.1$, and finally $t'=-0.25$. This behavior indicates that $t'$ controls the temperature scales where $T\sigma_{xy}$ approaches a constant or diverges positively or negatively, while $T\sigma_{xx}$ remains indifferent to $t'$ down to low temperatures. 

\begin{figure}
    \centering
    \includegraphics[width=0.8\textwidth]{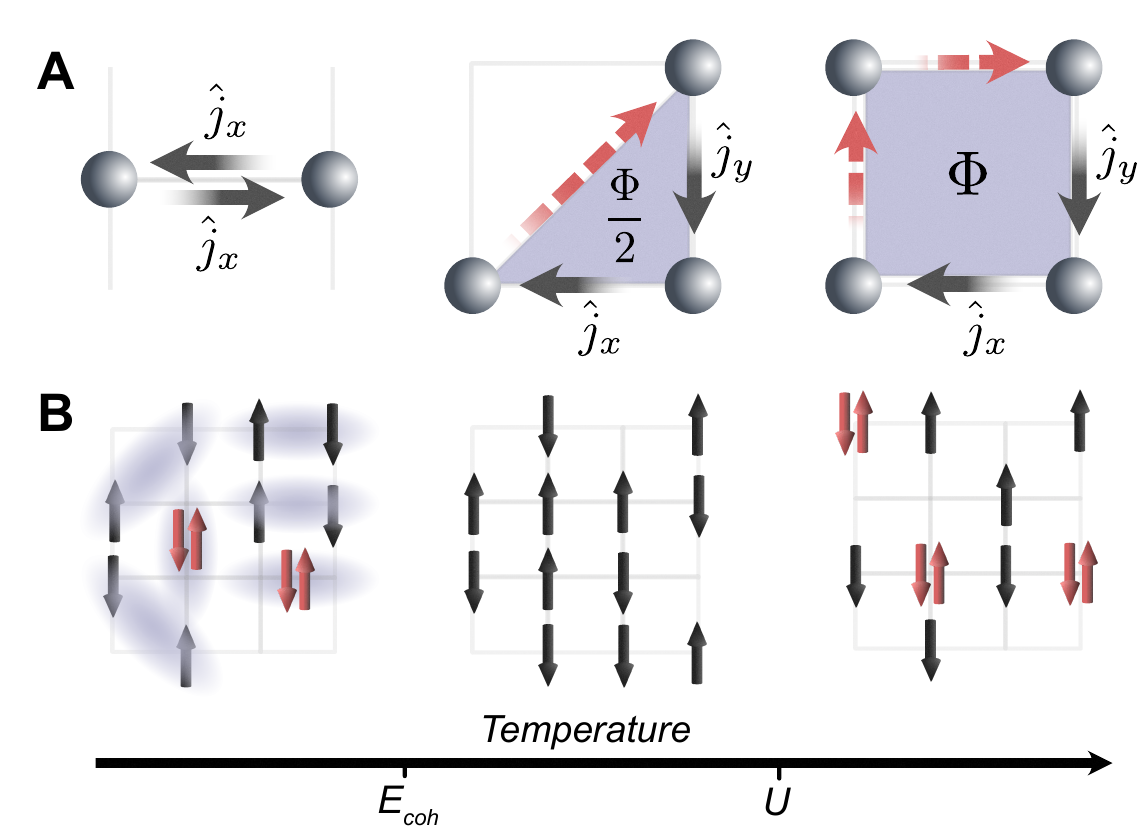}
    \caption{ \textbf{Conceptual sketches of longitudinal versus transverse transport and the semi-classical to quantum-coherent crossover. }
        (\textbf{A}) The conductivity $\langle\sigma_{\alpha\beta}\rangle\propto\mathrm{Tr}\left[e^{-\beta H} \hat{j}_\alpha \hat{j}_\beta\right]$ must have the same initial and final state in order for the trace to be nonzero. The longitudinal conductivity when $\alpha=\beta$ achieves this without intermediate virtual hopping processes (depicted as red dashed arrows), in contrast to the transverse conductivity when $\alpha\ne\beta$. With nonzero $t'$, triangular plaquette loops appear with an magnetic flux of $\Phi/2$. In the case of $t'=0$, two virtual hopping processes are needed, enclosing a flux of $\Phi$. 
        (\textbf{B}) At high temperature, the states $\{\uparrow, \downarrow, \uparrow\downarrow,\_ \}$ are equally thermally weighted resulting in a random distribution across sites. As temperature is lowered below $U$, the double occupancies are projected out due to strong electron repulsion. The formation of singlets (fuzzy purple ellipses) is accompanied by double occupancies as $T$ crosses over to the quantum-coherent regime.
    }
    \label{fig:schematic}
\end{figure}

The temperature at which $T\sigma_{xy}$ diverges coincides with the minimum in the average double occupancy $\langle d \rangle=\langle n_\uparrow n_\downarrow \rangle$, shown in the inset of Fig. \ref{fig:Tsigma}. Physically, the resurgence of $\langle d \rangle$ as temperature is lowered corresponds to a crossover from a semi-classical regime, where double occupancy is thermodynamically suppressed by Coulomb repulsion, to a quantum-coherent many-body regime, where fluctuations in double occupancy (doublons) stabilize singlet formation in the ground state, as illustrated in Fig.~\ref{fig:schematic}B. The temperature where $\langle d\rangle$ reaches its minimum provides an empirical proxy $E_{\text{coh}}$ for this crossover scale, and it tracks the same temperature dependence on $t'$ observed in the Hall conductivity.

To isolate the origin of this $t'$-dependence, we perform a high-temperature expansion (Supplementary Text)
\[
\sigma_{xy}
=
\frac{C_1}{T}
+
\frac{C_2}{T^2}
+
\frac{C_3}{T^3}
+\cdots,
\]
where coefficients $C_i$ measure enclosed hopping loops with magnetic flux. Physically, this expansion highlights the different requirements of the two transport channels. While longitudinal conductivity measures charge transport between neighboring sites, Hall conductivity requires the coherent accumulation of Aharonov-Bohm phases around closed loops, as illustrated in Fig.~\ref{fig:schematic}A. As a result, the Hall response acts as a sensitive detector for the onset of quantum coherence, tracking the temperature scale where multi-site correlation functions emerge from the high-temperature incoherent background. 

In the strong-coupling regime $(U\gg T)$, the first three coefficients can be expressed as
\begin{align}
    C_1 &= a_1\frac{t^2 t'}{UE_\text{coh}} + a_2 \frac{t^4}{U^2 E_\text{coh}} + \cdots \label{eq:c1}\\
    C_2 &= b_1\frac{t^2t'}{E_\text{coh}} + b_2\frac{t^4}{UE_\text{coh}}+\cdots\label{eq:c2} \\
    C_3 &= c_1 t^2t' + c_2\frac{t^4}{E_\text{coh}} + c_3\frac{t^4}{U}+\cdots.\label{eq:c3}
\end{align}
The key implication of Eqs.~(\ref{eq:c1})-(\ref{eq:c3}) is that the Hall conductivity depends not only on the sign of $t'$, which determines the orientation and amplitude of the lowest-order flux-enclosing loops, but also on the coherence scale $E_\mathrm{coh}$, which controls the onset of many-body quantum coherence. The DQMC results in Fig.~\ref{fig:Tsigma} indicate that $E_\mathrm{coh}$ is strongly renormalized by $t'$; for $t'>0$, deviations from the high-temperature behavior occur at temperatures of order $t$, whereas for $t'<0$, they occur at lower temperatures approaching the magnetic exchange scale $J$. This hierarchy is also reflected in Fig.~\ref{fig:spectralRH}, where the low-energy spectral weight either follows or departs from the underlying Fermi-surface geometry depending on the sign of $t'$.

The appearance of integer exponents in Fig.~\ref{fig:cot_theta_H} can be understood within this framework. From the Lehmann representation, the Hall conductivity admits an expansion of
\begin{align*}
    \sigma_{xy}\sim
    \begin{cases}
        T^{-1} & t'<0\\
        T^{-3} & t'\approx 0\\
        T^{-2} & t'>0
    \end{cases}
\end{align*}
with prefactors determined by the corresponding loop amplitudes. Consequently, whenever one coefficient dominates over a finite temperature range, the Hall conductivity naturally exhibits a simple power law with an integer exponent. Since $\sigma_{xx}\sim 1/T$, the Hall angle inherits the same hierarchy of integer powers. The DQMC results therefore suggest that different values of $t'$ select different dominant terms in the expansion.

Throughout the strange-metal regime, the temperature dependence of the Hall angle is dominated by the transverse response. The observed evolution from $\cot(\theta_H)\sim T^2$ near $t'\approx 0$, to 
$\cot(\theta_H)\sim T$ for $t'>0$, and $\cot(\theta_H)\sim$ constant for $t'<0$, therefore reflects changes in the dominant flux-enclosing hopping processes rather than changes in a transport scattering rate. In this picture, the Hall angle directly tracks the emergence of quantum-coherent loop motion.

The central result of this work is that the Hall response distinguishes between semi-classical and quantum-coherent transport in a manner that is largely invisible in the longitudinal channel. While the resistivity remains robustly linear in temperature across all parameter sets studied, the Hall conductivity displays pronounced sensitivity to particle-hole asymmetry, Fermi-surface geometry, and correlation-induced spectral-weight redistribution. Consequently, the inverse Hall angle exhibits several distinct temperature scalings rather than a universal quadratic form.
The observation of simple integer power laws in the Hall angle is particularly significant because it points to an underlying expansion in inverse temperature, rather than to a continuum of critical exponents or phenomenological scattering rates. 

The onset of these different scalings coincides with the coherence scale $E_\mathrm{coh}$ extracted independently from the temperature dependence of double occupancy. This correspondence suggests that Hall transport acts as a direct probe of the emergence of many-body quantum coherence. In contrast to the resistivity, which primarily reflects the existence of thermally active charge carriers, the Hall response is governed by coherent flux-enclosing hopping processes and is therefore sensitive to the formation of correlated electronic states.

More broadly, these results suggest that Hall transport may provide a universal window into coherence formation in strongly correlated matter. Whether in cuprates, nickelates~\cite{li_superconducting_2020}, moir\'e materials~\cite{lyu_strange_2021,cao_strange_2020,xia_bandwidth-tuned_2026}, or other systems exhibiting $T$-linear resistivity, the transverse response may contain information about emergent energy scales that is primarily absent from the longitudinal channel. Understanding this connection may offer a route toward a microscopic classification of strange metals based not on their resistivity, but on how quantum coherence first emerges within them.



\clearpage 

%
\bibliography{references} 
\bibliographystyle{sciencemag}

%
%
%
%
%
%


\section*{Acknowledgments}
We acknowledge helpful discussions with W.~O.~Wang, J.~K.~Ding, B.~Moritz, R.~Jin, E.~Tulipman, S.~Kivelson, and A.~Auerbach.

\paragraph*{Funding:}
This work was supported by the U.S. Department of Energy (DOE), Office of Basic Energy Sciences,
Division of Materials Sciences and Engineering. 
Computational work was performed on the Sherlock cluster at Stanford University and on resources of the National Energy Research Scientific Computing Center (NERSC), a Department of Energy Office of Science User Facility, using NERSC award ERCAP0034969. E.Z.Z. acknowledges support from the Geballe Laboratory of Advanced Materials Postdoctoral Fellowship. 

\paragraph*{Author contributions:}
E.Z.Z. and T.P.D conceived the study and wrote the manuscript. E.Z.Z. performed numerical simulations and conducted data analysis and interpretation. T.P.D. performed the high-temperature expansion. 

\paragraph*{Competing interests:}
``There are no competing interests to declare.''

\paragraph*{Data, code and materials availability:}
The aggregated numerical data and analysis scripts are available at 10.5281/zenodo.20622235. Raw simulation
data that support the findings of this study are stored
on the Sherlock cluster at Stanford University and are
available from the corresponding author upon reasonable
request.


\subsection*{Supplementary materials}
Methods\\
Supplementary Text\\
Figs. S1 to S3\\
References \textit{(35-\arabic{enumiv})}\\ 


\newpage


\renewcommand{\thefigure}{S\arabic{figure}}
\renewcommand{\thetable}{S\arabic{table}}
\renewcommand{\theequation}{S\arabic{equation}}
\renewcommand{\thepage}{S\arabic{page}}
\setcounter{figure}{0}
\setcounter{table}{0}
\setcounter{equation}{0}
\setcounter{page}{1} 


\begin{center}
\section*{Supplementary Materials for\\ \scititle}

Emily~Z.~Zhang,
Thomas~P.~Devereaux$^{\ast}$\\ 
\small$^\ast$Corresponding author. Email: tpd@stanford.edu\\
\end{center}

\subsubsection*{This PDF file includes:}
Methods\\
Supplementary Text\\
Figures S1 to S4\\

\newpage


\subsection*{Methods}
\subsubsection*{Simulation Parameters}
\begin{figure} 
	\centering
	\includegraphics[width=0.6\textwidth]{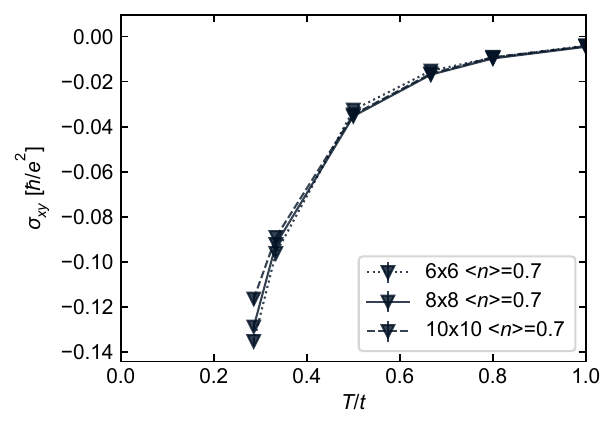}

	\caption{\textbf{Finite size testing for $\sigma_{xy}$.}
	   Parameters used: $U/t=6$, $\langle n \rangle = 0.7$, $t'/t=-0.1$. The magnetic field strength is $\Phi/\Phi_0=2/36$ for 6x6, $\Phi/\Phi_0=4/64$ for 8x8, and $\Phi/\Phi_0=6/100$ for 10x10.  }
	\label{fig:finite_size} 
\end{figure}

\begin{figure} 
	\centering
	\includegraphics[width=1.0\textwidth]{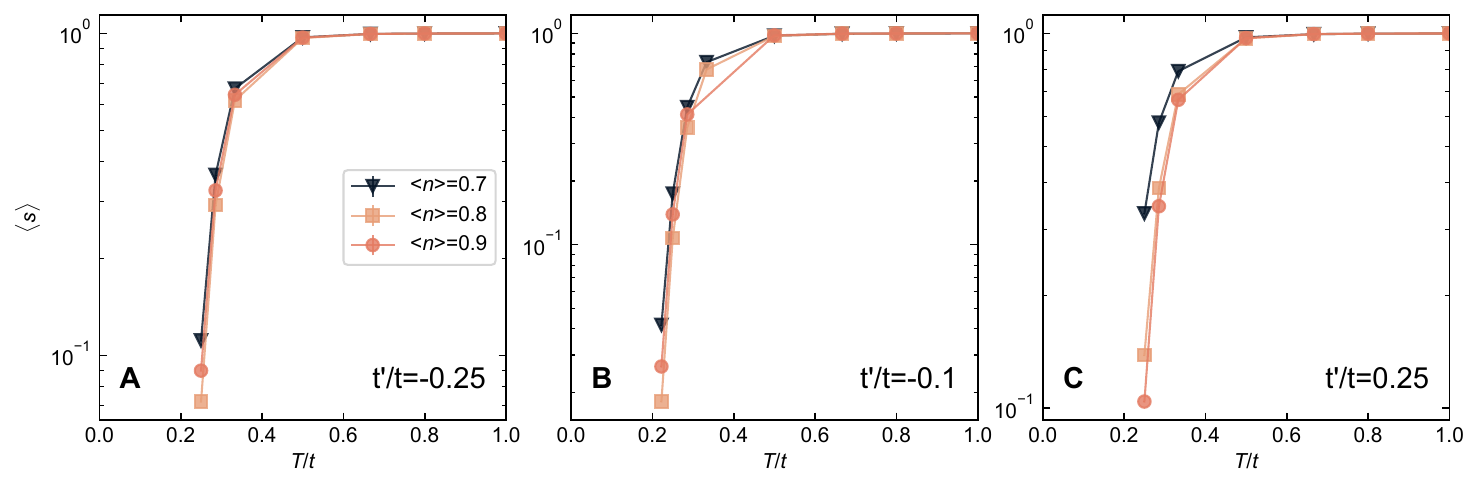}

	\caption{\textbf{Average sign $\langle s \rangle$ as a function of temperature, $t'$, and doping.}
		Parameters used: $U/t=6$, $B=0.0625\Phi_0/a^2$.  }
	\label{fig:sign} 
\end{figure}

Determinant quantum Monte Carlo (DQMC) was used to simulate the 2D single-band Hubbard-Hofstader model in the grand canonical ensemble. All simulations were performed on 8x8 square lattice clusters (see finite size testing in Fig.~\ref{fig:finite_size}) with modified periodic boundary conditions for the transport measurements~\cite{assaad_depleted_2002}, and twisted boundary conditions for the spectral measurements~\cite{wang_probing_2025}. We ran $4\times 10^4$ warm-up sweeps and between $4\times 10^5$ and $1\times 10^6$ measurement sweeps through the auxiliary field. To mitigate the sign problem for our chosen parameters (see Fig.~\ref{fig:sign}), at least 600 independently seeded Markov chains were used for the transport measurements, and 1000 were used for the spectral measurements. For all parameter values, we used a small imaginary time discretization of $\Delta\tau\le 0.05/t$ and fixed the number of imaginary time slices to $L=\beta/\Delta\tau$ to reduce Trotter error. For the unequal time measurements, we took every 2 measurements per auxiliary field sweep. For all doping, we tuned the chemical potential such that $|\langle n_{\text{target}} \rangle-1|\le 1\times 10^{-4}$. 

\subsubsection*{Analytic Continuation for Transport}
To compute the current-current correlators, we employed maximum entropy (MaxEnt) as an analytic continuation algorithm to invert the equation 
\begin{equation}
    \chi_{\alpha\beta}(\tau)=\frac{1}{V}\langle J_\alpha (\tau)J_\beta(0)\rangle=\int_{-\infty}^{\infty}
    \mathrm{d}(\hbar\omega)~\frac{e^{-\tau\omega\hbar}}{1-e^{-\beta\omega\hbar}} \frac{-\text{Im}[\chi_{\alpha\beta}(\omega)]}{\pi}.
\end{equation}
For $\alpha=\beta$, the longitudinal conductivity can then be calculated by 
\begin{equation}
    \text{Re}~\sigma_{xx}(\omega)=-\frac{\text{Im}[\chi_{xx}(\omega)]}{\omega}.
\end{equation}
For the transverse component, however, one cannot directly use MaxEnt to retrieve $\chi_{xy}(\omega)$ because the spectral function is not guaranteed to be positive definite. We therefore use a subtraction method~\cite{ding_intrinsic_2025}, where we perform MaxEnt on composite object 
\begin{equation}
    \chi_{\text{comp.}}(\tau)=\chi_{xx}(\tau)-i\chi_{xy}(\tau),
\end{equation}
and subtract off the longitudinal piece to obtain $\text{Re}[\chi_{xy}(\omega)]$ and subsequently 
\begin{equation}
    \text{Im}[\sigma_{xy}(\omega)]=\frac{\text{Re}[\chi_{xy}(\omega)]}{\omega}.
\end{equation} 
The DC component of $\sigma_{xy}$ can then finally be obtained using a Kramer's Kronig transformation. Since the composite object is Hermitian, the spectral function of $\chi_{\text{comp.}}$ is guaranteed to be postive definite, allowing us to use MaxEnt. We imposed additional constraints on the MaxEnt algorithm to ensure the correct symmetries were obeyed, which will be described in detail in a future paper with Rebekah Jin. We used a flat model function and chose hyper-parameter $\alpha$ using the "BT" method in our MaxEnt fits~\cite{bergeron_algorithms_2016}. The mean and standard error of the analytic continuation results were obtained from bootstrap resamplings of at least 100 samples. All transport figures were computed using MaxEnt unless stated otherwise. 

\subsubsection*{Proxy for Spectral Function and Conductivity}
\begin{figure} 
	\centering
	\includegraphics[width=0.9\textwidth]{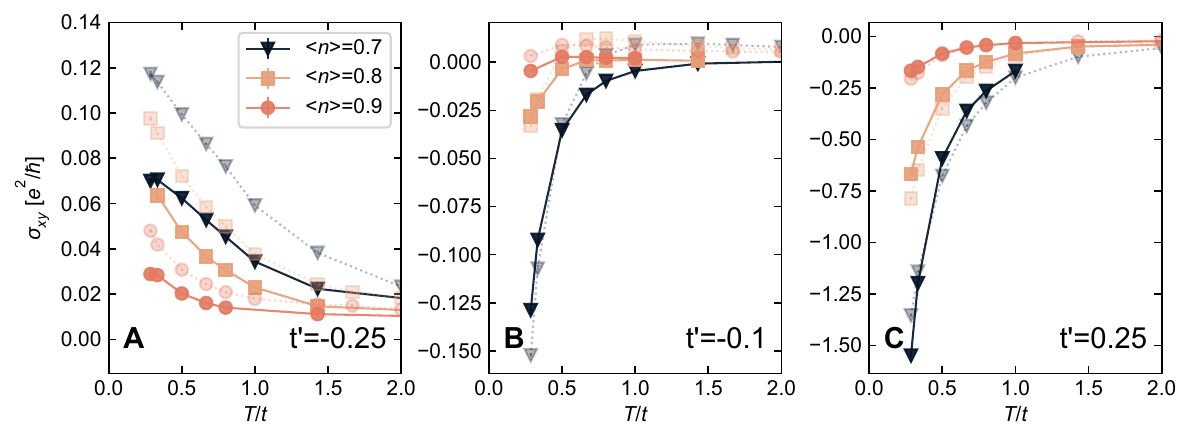}

	\caption{\textbf{Comparison of proxy and MaxEnt for $\sigma_{xy}$.}
        The MaxEnt results are depicted as opaque solid lines, and the proxy results are depicted as transparent dashed lines. Parameters used: $U/t=6$, $B=0.0625\Phi_0/a^2$.}
	\label{fig:ACproxy} 
\end{figure}

Proxies are an alternative way to estimate dynamical quantities without needing to use analytic continuation. For the transverse Hall conductivity, we estimated the DC component using the Hall coefficient at the first Matsubara frequency (M1) as a proxy~\cite{wang_numerical_2021-1,ding_intrinsic_2025} defined as 
\begin{equation}
R_{\mathrm{H}}^{\mathrm{M1}}(i\omega_n) = \frac{1}{B} \frac{\chi_{xy}(i\omega_n) \omega_n T}{(\chi_{xx}(i\omega_n) - \chi_{xx}(0))^2 + \chi_{xy}(i\omega_n)^2}  \label{eq:proxy-m1}
\end{equation}
where $\chi_{xx}$ and $\chi_{xy}$ are the current correlators as a function of Matsubara frequency $\omega_n$. $\chi_{xx}(0)$ can be obtained from the analytic continuation method described above. To estimate the DC conductivity at finite temperature, we take $R_{\mathrm{H}}^{\mathrm{M1}}(i\omega_n)$ at the lowest nonzero Matsubara frequency $\omega_1 = 2\pi/\beta$ as a proxy for $R_{\mathrm{H}}^{\mathrm{DC}}$. This approximation is valid as long as $R_H$ is well-behaved in imaginary time, and $\omega_n$ does not change rapidly for small $n$.
Additionally, we used $\beta G_\mathbf{k}(\tau=\beta/2)$ as a proxy to estimate the spectral function $A(\mathbf{k},\omega=0)$~\cite{wang_probing_2025}. Both proxies have shown good qualitative agreement with their respective analytically continued quantities~\cite{wang_numerical_2021-1,wang_probing_2025}. The comparison for the Hall conductivity is shown in Fig.~\ref{fig:ACproxy}. 


\subsection*{Supplementary Text}

\subsubsection*{High Temperature Expansions of $\sigma_{xx}$ and $\sigma_{xy}$}
In Lehmann representation, the DC longitudinal conductivity $\sigma_{xx}$ and Hall conductivity $\sigma_{xy}$ are given by 
\begin{align}
    \mathrm{Re}\,\sigma_{xx}&=\frac{\pi\hbar}{ZVT}\sum_{n,m} e^{-\beta E_n} |\bra{n}\hat j_x\ket{m}|^2\delta( E_n - E_m)\text{ and} \label{eq:sigma_xx}\\
    \mathrm{Re}\,\sigma_{xy}
&=
\frac{\hbar}{V}
\sum_{m\neq n}
\frac{
e^{-\beta E_n}-e^{-\beta E_m}
}{
(E_m-E_n)^2
}
\,
\mathrm{Im}
\left\{
\bra{n}\hat j_x\ket{m}\bra{m}\hat j_y\ket{n}
\right\},\label{eq:sigma_xy}
\end{align}
where $\hbar$ is Planck's constant, $Z$ is the partition function, $V$ is the volume, $T$ is the temperature, $\beta=1/k_BT$, indices $n$ and $m$ sum over the eigenstates of the Hamiltonian, and $\hat{j}_{x,y}$ are the current density operators in the $x$ and $y$ directions. 

For the longitudinal conductivity Eq.~(\ref{eq:sigma_xx}), the explicit temperature dependence enters through the prefactor $1/T$ and the Boltzmann weights
$e^{-\beta E_n}$. When the many-body excitations that dominate transport lie within an energy window $E_{\mathrm{coh}}\lesssim k_BT$, where $E_{\mathrm{coh}}$ denotes the emergent many-body scale governing transport, the Boltzmann factors vary only weakly across the thermally accessible states. Provided that the current matrix elements do not acquire singular temperature dependence, the sum in Eq.~(\ref{eq:sigma_xx}) remains approximately temperature independent, yielding
\[
\sigma_{xx}\sim \frac{1}{T}.
\]
This scaling should not be interpreted as a conventional high-temperature expansion. The relevant criterion is not $k_BT\gg t$, but rather $k_BT\gg E_\mathrm{coh}$. In the doped Hubbard model, strong correlations can suppress 
$E_\mathrm{coh}$ far below the bare electronic bandwidth. Consequently, the transport-weighted many-body spectrum remains effectively thermally populated over an extended temperature range, and the explicit $1/T$ prefactor in Eq.~(\ref{eq:sigma_xx}) controls the leading temperature dependence of $\sigma_{xx}$. 

The temperature dependence of the transverse conductivity Eq.~(\ref{eq:sigma_xy}) is substantially more constrained by symmetry. Unlike the longitudinal response, the Hall conductivity is odd under time reversal and therefore vanishes unless time-reversal symmetry is broken by an external magnetic field. In lattice systems, a finite Hall response arises from virtual hopping processes that encircle magnetic flux and acquire phase information from the underlying electronic motion, depicted in Fig.~\ref{fig:schematic}A in the main text. Consequently, $\sigma_{xy}$ probes not only the availability of thermally populated states, but also the geometric structure of the transport pathways connecting them. This additional sensitivity makes the Hall conductivity a particularly valuable probe of emergent energy scales and correlation effects. As a result, $\sigma_{xy}$ generally reveals additional low-energy scales that may remain hidden in the longitudinal conductivity.

Defining
\begin{equation*}
\Delta_{mn}=E_m-E_n,
\end{equation*}
the thermal factor in Eq.~(\ref{eq:sigma_xx}) admits the expansion
\begin{equation}
\frac{
e^{-\beta E_n}-e^{-\beta E_m}
}{
(E_m-E_n)^2
}
=
\frac{\beta}{\Delta_{mn}}
-\frac{\beta^2}{2}
+\frac{\beta^3}{6}\Delta_{mn}
+\mathcal O(\beta^4).
\end{equation}
Substituting this series into Eq.~(\ref{eq:sigma_xx}) generates a systematic high-temperature expansion of the Hall conductivity
\begin{equation}
\mathrm{Re}\,\sigma_{xy}
=
\beta C_1+\beta^2 C_2+\beta^3 C_3+O(\beta^4),
\end{equation}
where the coefficients $C_n$ are temperature-independent
moments of the many-body spectrum and current matrix elements.
Although derived formally as an expansion about $\beta=0$,
the leading terms remain informative whenever the
transport-relevant excitation energies satisfy
$\Delta_{mn}\lesssim k_BT$.

Details about how these coefficients are obtained will appear in a forthcoming work. To motivate Eqs.~(\ref{eq:c1})--(\ref{eq:c3}), it is useful to consider the strong-coupling regime $(U\gg k_BT)$, where transport may be organized in terms of virtual hopping processes. The Hall response requires oriented hopping loops that distinguish clockwise and counterclockwise motion in the presence of magnetic flux. Consequently, the leading contributions to $C_n$ arise from the lowest-order closed paths that enclose a finite area.

For finite $t'$, the smallest such loops involve two nearest-neighbor hoppings and one next-nearest-neighbor hopping, yielding amplitudes proportional to $t^2t'$. In the strongly correlated regime, these processes may additionally involve virtual doublon-holon excitations with energy cost $U$, producing factors of $1/U$. The resulting contributions generate the leading terms proportional to $\frac{t^2t'}{U E_{\rm coh}}$ in $C_1$ and $\frac{t^2t'}{E_{\rm coh}}$ in $C_2$.

Higher-order contributions arise from longer hopping sequences. In particular, four-step processes produce terms proportional to $t^4$. Depending on whether these paths involve virtual charge excitations, their amplitudes are controlled either by the interaction scale $U$ or the emergent coherence scale $E_{\rm coh}$, leading to terms of order $\frac{t^4}{U^2E_{\rm coh}}, \frac{t^4}{UE_{\rm coh}}, \frac{t^4}{E_{\rm coh}}$, and $\frac{t^4}{U}$, as summarized in Eqs.~(\ref{eq:c1})--(\ref{eq:c3}).

Since the low-energy coherence scale governing transport $E_\mathrm{coh}$ in the strong-coupling Hubbard model may be parametrically smaller than both $t$ and $U$, the nominally subleading terms in Eqs.~(\ref{eq:c1})--(\ref{eq:c3}) may dominate over an extended temperature range.

The coefficients $a_i, b_i,$ and $c_i$ depend on microscopic details of the many-body spectrum and current matrix elements. Their explicit evaluation requires a systematic strong-coupling expansion of the Hall conductivity and will be presented elsewhere. Here we emphasize only the hierarchy of energy scales and hopping processes that controls the structure of Eqs.~(\ref{eq:c1})--(\ref{eq:c3}).


\begin{figure} 
	\centering
	\includegraphics[width=0.9\textwidth]{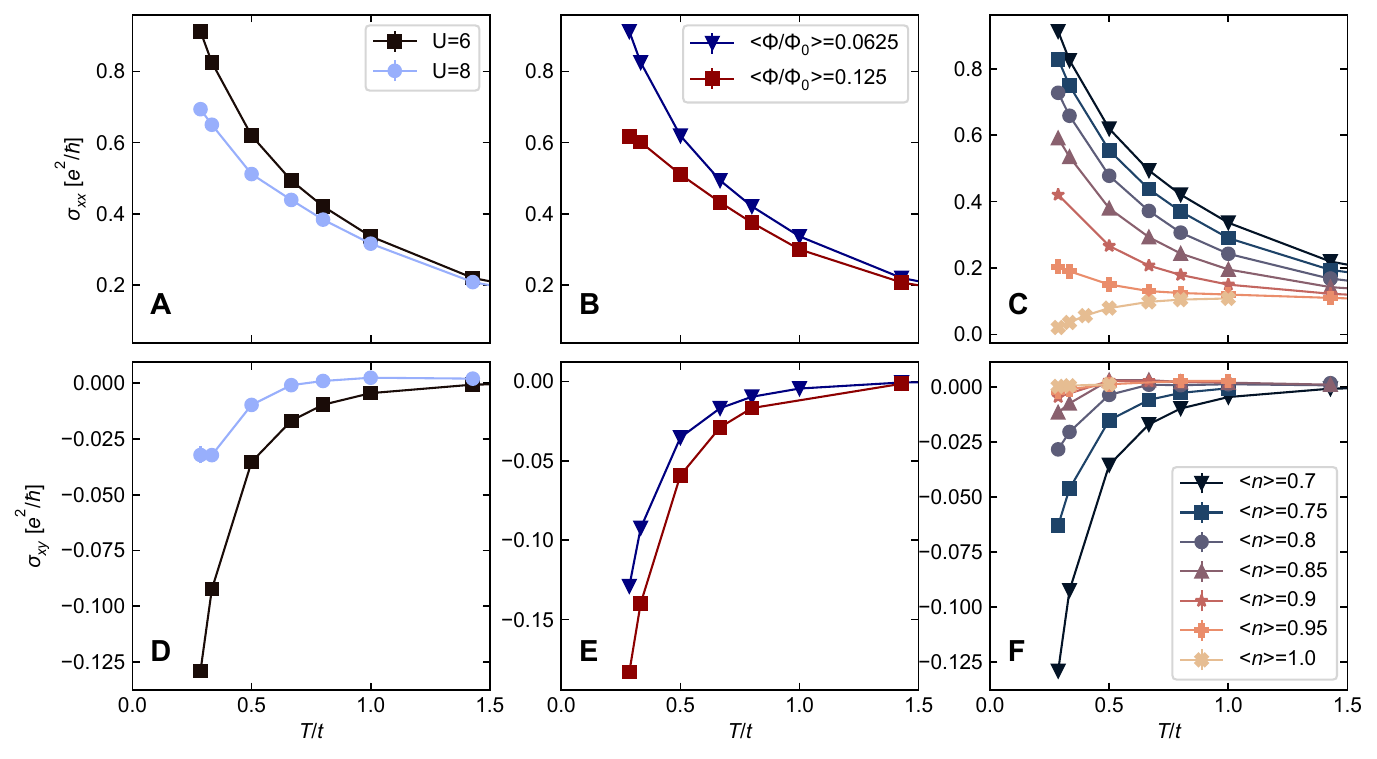}

	\caption{\textbf{$U$, $B$, and doping dependence of longitudinal and transverse conductivity.}
        (\textbf{A}) and (\textbf{D}) were calculated with $\langle n \rangle = 0.7$ and $B=0.0625\Phi_0/a^2$. (\textbf{B}) and (\textbf{E}) were calculated with $U/t=6$ and $\langle n \rangle = 0.7$. (\textbf{C}) and (\textbf{F}) were calculated with $U/t=6$ and $B=0.0625\Phi_0/a^2$. All plots were calculated with $t'/t=-0.1$. }
	\label{fig:U_nflux_doping_sweep} 
\end{figure}




\end{document}